\documentstyle[12pt,aasms4]{article}
\begin{document}

\title{Gamma Radiation from PSR B1055$-$52}

\author{D. J. Thompson\altaffilmark{1,2},  M. Bailes\altaffilmark{3}, D. L. Bertsch\altaffilmark{1}, J. Cordes\altaffilmark{4},
 N. D'Amico\altaffilmark{5}, J. A. Esposito\altaffilmark{1,6},  J. Finley\altaffilmark{7},
R. C. Hartman\altaffilmark{1}, W. Hermsen\altaffilmark{8},  
G. Kanbach\altaffilmark{9},  V. M. Kaspi\altaffilmark{10}, D. A. Kniffen\altaffilmark{11},
 L.~Kuiper\altaffilmark{8}, Y. C. Lin\altaffilmark{12}, A. Lyne\altaffilmark{13},
 R. Manchester\altaffilmark{14}, S.M. Matz\altaffilmark{15}, H.~A.~Mayer-Hasselwander\altaffilmark{9}, P. F. Michelson\altaffilmark{12},  P. L. Nolan\altaffilmark{12},
H. \"Ogelman\altaffilmark{16}, M. Pohl\altaffilmark{17}, P.~V.~Ramanamurthy\altaffilmark{1,18}, P. Sreekumar\altaffilmark{2}, O. Reimer\altaffilmark{9}, J.H. Taylor\altaffilmark{19},
M. Ulmer\altaffilmark{15}}

\altaffiltext{1}{Code 661, Laboratory for High Energy Astrophysics, NASA Goddard Space Flight Center, Greenbelt, MD 20771 USA}
\altaffiltext{2}{djt@egret.gsfc.nasa.gov}
\altaffiltext{3}{Astrophys. and Supercomputing, Mail No. 31, Swinburne Univ. of Technology, PO Box 218, Hawthorn  Victoria 3122, Australia}
\altaffiltext{4}{Physics Dept., Cornell University, Ithaca, NY }
\altaffiltext{5}{Osservatorio Astronomico di Bologna, via Zamboni 33, I-40126 Bologna}
\altaffiltext{6}{USRA Research Associate}
\altaffiltext{7}{Physics Dept., Purdue University, Lafayette, IN }
\altaffiltext{8}{SRON/Utrecht, Sorbonnelaan 2, 3584 CA Utrecht, The Netherlands}
\altaffiltext{9}{Max-Planck-Institut f\"ur Extraterrestrische Physik,
Giessenbachstr D-85748 Garching FRG}
\altaffiltext{10}{Physics Dept., MIT, Cambridge, MA USA}
\altaffiltext{11}{Department of Physics, Hampden-Sydney College,
  Hampden-Sydney, VA 23943 USA}
\altaffiltext{12}{W.W. Hansen Experimental Physics Laboratory and
Department  of Physics, Stanford University, Stanford CA
94305 USA}
\altaffiltext{13}{Nuffield Radio Astronomy Laboratories, Jodrell Bank, University of Manchester, Macclesfield, Cheshire SK11 9DL, UK }
\altaffiltext{14}{ATNF, CSIRO, Australia }
\altaffiltext{15}{Physics Dept., Northwestern University, Evanston, IL }
\altaffiltext{16}{Physics Dept., University of Wisconsin, Madison, WI }
\altaffiltext{17}{Danish Space Research Institute, 2100 Copenhagen O, Denmark}
\altaffiltext{18}{National Academy of Sciences-National Research Council Senior Research Associate, Present address: S.T.E. Laboratory, Nagoya University, Nagoya, Japan}
\altaffiltext{19}{Physics Dept., Princeton University, Princeton, NJ}

\begin{abstract}
The telescopes on the 
Compton Gamma Ray Observatory (CGRO) have observed PSR B1055$-$52 a number of times between 1991 and 1998.  From these data, a more detailed picture of the gamma radiation from this source has been developed, showing several characteristics which distinguish this pulsar: the light curve is complex; there is no detectable unpulsed emission; the energy spectrum is flat, with no evidence of a sharp high-energy cutoff up to $>$4 GeV.  Comparisons of the gamma-ray data with observations at longer wavelengths show that no two of the known gamma-ray pulsars have quite the same characteristics; this diversity makes interpretation in terms of theoretical models difficult.
\end{abstract}


\section{Introduction}

The Compton Gamma Ray Observatory (CGRO) telescopes have  detected pulsed gamma radiation from at least seven spin-powered pulsars:  Crab; Vela; Geminga; PSR B1509$-$58; PSR B1706$-$44; PSR B1951+32; and PSR B1055$-$52, with some evidence for an eighth, PSR B0656+14.  For a summary of CGRO pulsar results, see Ulmer (1994) and Thompson et al.(1997).  Upper limits have been calculated for selected samples of radio pulsars (Thompson et al. 1994; Fierro et al. 1995; Carrami\~nana et al. 1995, Schroeder et al. 1995) and for all cataloged pulsars (Nel et al. 1996).     

The present work is a detailed analysis of the gamma-ray observations of PSR B1055$-$52, based on repeated observations during 1991$-$1998 which have nearly tripled the source exposure time compared to the discovery data (Fierro et al. 1993).  The gamma-ray observations of this and other pulsars are shown in a multiwavelength context. Comparison of the multiwavelength properties of pulsars is important in attempting to construct models for these objects.

Using ROSAT, \"Ogelman and Finley (1993) found pulsed X-rays from PSR B1055$-$52, with a pulse that changed both shape and phase at photon energy about 0.5 keV.  The X-ray energy spectrum requires at least two components, one thought to be emission from the hot neutron star surface and the other likely to be from the pulsar magnetosphere (see also Greiveldinger et al. 1996, and Wang et al. 1998).

Mignani, Caraveo, and Bignami (1997) have found evidence based on positional coincidence for an optical counterpart of PSR B1055$-$52 using the Hubble Space Telescope Faint Object Camera.  In the absence of a fast photometer on HST and the presence of a nearby bright star, finding optical pulsations will be difficult, as noted by the authors.

\section{Radio Observations} 

The basic pulsar parameters (Taylor, Manchester, \& Lyne 1993), derived from radio observations, are shown in Table 1.

\placetable{1}

PSR B1055$-$52 was on the list of nearly 300 pulsars monitored regularly by radio astronomers to assist gamma-ray telescopes on the Compton Observatory (Kaspi, 1994; Arzoumanian, et al. 1994; Johnston et al. 1995; D'Amico et al. 1996 ).  High-energy gamma-ray data are sparse; weak, short-period gamma-ray pulsars are detectable only if the timing parameters are determined independently of the gamma-ray data. In the case of PSR B1055$-$52, this monitoring, carried out at Parkes, has continued.   The pulsar exhibits considerable timing noise.  For this reason, the timing solutions used for the gamma-ray analysis were developed piecewise over time intervals for which the pulse phase could be adequately modeled using only a simple spin-down law in terms of $\nu$,  $\dot\nu$, and $\ddot\nu$.  Table 2 lists the solutions relevant to the Compton Observatory viewings, given in terms of frequency $\nu$ and its derivatives instead of period, and valid at time $T_0$.  These timing solutions are from the database maintained at Princeton University (anonymous FTP: ftp://puppsr.princeton.edu/gro) with the addition of recent timing solutions from Parkes (http://wwwatnf.atnf.csiro.au/research/pulsar/psr/archive).

\placetable{2}

\section{Gamma-Ray Observations}

	All the telescopes on the Compton Observatory have pulsar timing capability. Time  in Universal Coordinated Time (UTC) is carried on board the spacecraft to an accuracy of better than 100 $\mu$s.  The conversion of gamma-ray arrival time at the location of the Compton Observatory to pulsar phase is carried out  using a modification of the TEMPO timing program (Taylor \& Weisberg 1989) and the Jet Propulsion Laboratory DE200 ephemeris.

        EGRET is the high-energy gamma-ray telescope on CGRO (Thompson et al.  1993), operating from about 30 MeV to over 20
GeV.    The field of view mapped by EGRET extends to more than 30$^\circ$ from the instrument axis. PSR B1055$-$52 was within 30$^\circ$ of the telescope axis during 24 of the CGRO viewing intervals to date.  No additional EGRET observations of this pulsar are scheduled.  PSR B1055$-$52 is not a particularly bright source compared to many others seen by EGRET (cf. the second EGRET catalog, Thompson et al. 1995). Its gamma-ray count rate is low, about 4 photons (E$>$100 MeV) per day when the source is within 10$^\circ$ of the EGRET axis. 
Data processing for EGRET relies on two principal methods -- timing analysis and spatial analysis. The spatial analysis compares the observed gamma-ray map to that expected from a model of the diffuse Galactic and extragalactic radiation (Hunter et al. 1997; Sreekumar et al. 1998). Source location and flux as a function of energy are determined using a maximum likelihood method (Mattox et al. 1996).  The timing and spatial approaches can also be combined to produce phase-resolved maps and energy spectra. 

COMPTEL, the Imaging Compton Telescope, is another of the Compton Observatory telescopes, operating in the energy range 0.75--30 MeV (Sch\"onfelder et al 1993). Like EGRET, COMPTEL uses both spatial and timing analysis, and because the COMPTEL field of view is larger than EGRET's and the two telescopes are co-aligned on the spacecraft, PSR B1055$-$52 was viewed by COMPTEL at the same times as by EGRET.

OSSE, the Oriented Scintillation Spectrometer Experiment, is a third of the CGRO telescopes, operating in the energy range 0.05 -- 10 MeV (Johnson et al. 1993).  Like EGRET and COMPTEL, OSSE uses both spatial and timing analysis, with the spatial analysis coming from an 
on-source/off-source analysis.  Due to its smaller field of view, OSSE observes individual targets.  PSR B1055$-$52 was observed by OSSE and simultaneously by EGRET and COMPTEL  1997 Sept. 2--9 and 1997 Sept. 23--Oct. 7. 

\section{Results}

\subsection{Light Curves}

Fig. 1 shows the EGRET light curve for PSR B1055$-$52 combining data from all 24 viewings, derived in two different ways: (Top) The gamma-ray selection was based on maximizing the significance of the pulsed signal in the light curve, as characterized by the (binned) $\chi^2$ or (unbinned) H-test value (De Jager, Swanepoel, \& Raubenheimer 1978).  The strongest signal was obtained for  energies above 240 MeV with gamma rays selected within 1.7$^{\arcdeg}$ of the known pulsar position.  As found by Ramanamurthy et al. (1995a) for PSR B1951+32, a fixed cone can produce an improved signal to background for relatively weak signals, because the fixed cone selects photons from the narrow component of the point spread function of the EGRET instrument (Thompson et al. 1993) for this energy range, eliminating the broad wings that contribute to the standard energy-dependent event selection. The optimization was done iteratively on angle and energy, involving about 20 trials. (Bottom) The alternate gamma-ray selection used  photons selected within an energy-dependent cone of radius
\begin{equation}
\theta \le 5\fdg85 {(E_\gamma/{100\phn {\rm MeV}})}^{-0.534},
\end{equation}

\noindent with respect to the pulsar position ($E_\gamma$ in MeV). This
choice represents the 67\% containment angular resolution of the EGRET
instrument (Thompson et al. 1993), including both narrow and broad components.  In this case the strongest signal was obtained for photon energies greater than 600 MeV.  For both light curves, the phase is the same, referenced to one of the two radio pulses that defines phase 0 (long-dash line).  The centroid of the second radio pulse is  at phase 0.43 (see, e.g. Biggs 1990), indicated in the figure by a short-dash line. 

The two gamma-ray light curves are not independent.  Most of the 146 photons in the lower light curve are also contained in the 328-photon upper curve.  Nevertheless, the similarity of the light curve shapes shows that the features are not the result of a particular selection technique. The light curve is different from those of the Crab, Vela, Geminga and PSR B1951+32, all of 
which show two narrow peaks separated by 0.4$-$0.5 in phase.  PSR B1055$-$52 more closely resembles PSR B1706$-$44 (Thompson et al. 1996) in having a broad phase range of emission.  Between phases 0.7 and 1.1, PSR B1055$-$52 shows evidence of two peaks.   The upper, higher-statistics light curve,  is reasonably well fit by two gaussians plus a constant term (reduced $\chi^2$ = 1.2, 23 degrees of freedom, probability 0.25) but not well by a single gaussian plus a constant (reduced $\chi^2$ = 2.2, 26 degrees of freedom, probability $<$ 0.001) or by a square pulse (reduced $\chi^2$ = 2.1, similar to the single gaussian).  Adding the second peak increases the F-test statistic from 20.7 to 43.7, a marked improvement.  The best fit is obtained with the following parameters: peak 1 phase   0.75, $\sigma_1$=0.02 (4.5 ms); peak 2 phase 0.95, $\sigma_2$=0.07 (14.2 ms).  In light of the limited statistics, details of the pulse shape cannot be considered well-defined. 
The significance of the small peak near phase 0.52 can be assessed by  calculating the Poisson probability of finding one bin out of 18 in the off-pulse region with 13 or more counts when the average in this phase region is 5.33 counts/bin.  The resulting 6\% probability indicates that none of the features in the 0.1--0.7 phase range is statistically significant.  

Timing analysis of the COMPTEL data produced no statistically-significant detection of pulsed emission in any of the bands 0.75--1, 1--3, 3--10, or 10--30 MeV.  Each of the bands and the summed COMPTEL light curve do, however, show a low-significance peak at phase $\sim$0.73 consistent in phase with the narrower of the two EGRET peaks.  Taking the EGRET pulse as a reference to define a preferred phase, the statistical significance of the peak in the summed COMPTEL light curve for a single trial is 3.5 $\sigma$, with a probability of chance occurrence less than 0.001.   Although the EGRET statistics do not warrant a detailed spectral analysis for the two pulses separately, it is noted that the narrow pulse does not appear above 2 GeV while the broad pulse extends to more than 4 GeV, suggesting that the narrow pulse may have a softer spectrum than the broad pulse, consistent with the peak seen by COMPTEL and the peak seen in hard X-rays.   The COMPTEL light curve is shown in Fig. 2, along with light curves from other wavelengths.  The vertical dashed line marks the reference radio peak, defining phase 0 in Fig. 1.

Timing analysis of the OSSE data produced no evidence of pulsed emission, even taking into account the constraints of assuming the same light curve shape at seen at the higher energies.  The 95\% confidence upper limit in the energy range 50--200 keV is  2.1  $\times$ 10$^{-4}$ ph cm$^{-2}$ s$^{-1}$MeV$^{-1}$.  The OSSE light curve is also shown in Fig. 2.

\subsection{Search for Unpulsed Emission and Source Variability}

In the region of sky mapped by EGRET around the pulsar, each photon's arrival time is converted to pulsar phase, whether or not this photon is likely to have come from the pulsar itself.  Phase-resolved maps of the sky are then constructed, the spatial analysis using maximum likelihood (Mattox et al. 1996) is used to assess the statistical significance of a source at the pulsar location.  The likelihood ratio test is used to determine the significance of
point sources. The likelihood ratio test statistic is $TS \equiv2(lnL_1-lnL_0)$, where
$lnL_1$ is the  log of the likelihood of the data if a point source is included
in the model, and $lnL_0$ is log of the likelihood of the data without
a point source.  For positive values of TS, the analysis gives the most likely gamma-ray flux of a source at the pulsar location.   The pulsar is detected with high significance between phases 0.7 and 1.1. Based on the summed map for all observations, the time-averaged flux above 100 MeV for this phase range is  (2.5 $\pm$ 0.2) $\times$ 10$^{-7}$ ph cm$^{-2}$ s$^{-1}$, and the statistical significance of the detection is 13.6$\sigma$. The flux is consistent with the value of (2.2 $\pm$ 0.4) $\times$ 10$^{-7}$ in the second EGRET catalog (Thompson et al. 1995). Analysis of the off-pulse phase range 0.1 to 0.7 yields an excess with a statistical significance of 1.9 $\sigma$, too small to claim a detection.  The upper limit (95\% confidence) is 1.2 $\times$ 10$^{-7}$ ph cm$^{-2}$ s$^{-1}$.     Above 100 MeV, any unpulsed component is therefore less than 50\% of the pulsed  emission. 

In a search for time variability, we examined the E$>$100 MeV
 observations of the PSR B1055$-$52 flux as a function of 
time, from 1991 $-$ 1997, for those 10 observations when the pulsar was within 20$\arcdeg$ of the EGRET axis, based on the maps of phase range 0.7 to 1.1.    As seen for the Crab, 
Geminga, and Vela pulsars (Ramanamurthy et al. 1995b) and PSR B1706$-$44 (Thompson et al. 1996), the high-energy gamma radiation from PSR B1055$-$52 appears to be steady.  The $\chi^2$ is 13.9 for 13 degrees of freedom.

\subsection{Energy Spectrum}

Because there is no substantial evidence for unpulsed emission (phase 0.1--0.7), the energy spectrum of the pulsar can be derived by analyzing the 0.7--1.1 phase map in each of 10 energy bands, using the maximum likelihood method as described above.  Including the few excess photons from the unpulsed region (less than 15\% increase in statistics) has no significant influence on the spectrum.  Nearby sources from the third EGRET catalog (Hartman et al. 1998) are included in the analysis, because the point spread functions for these sources overlap that of the pulsar, especially at the lower energies.  The excesses in each band are then compared to model spectra forward-folded through the EGRET energy response function, as described by Nolan et al. (1993).  Pulsed emission is detected from 70 MeV to more than 4 GeV.  The EGRET spectrum, shown in Fig. 3 as a phase-averaged photon number spectrum, can be represented by a power law

\begin{equation}
{dN\over{dE}} = (7.67 \pm 0.70) \times 10^{-11}{\left(E \over{541\: MeV}\right)}^{-1.73 \pm 0.08}  {\rm {photons \: cm}}^{-2} {\rm s}^{-1} {\rm MeV}^{-1},
\end{equation}

\noindent The reduced $\chi^2$ for this fit is 1.19.  Alternately, the spectrum can be fit as a broken power law, with a spectral break  at $\sim$1000 MeV, similar to the spectrum for PSR B1706$-$44 (Thompson et al. 1996)

 \begin{equation}
{dN\over{dE}} = (3.22 \pm 0.59) \times 10^{-11}{\left(E \over{1000\: MeV}\right)}^{\alpha}  {\rm {photons \: cm}}^{-2} {\rm s}^{-1} {\rm MeV}^{-1},
\end{equation}
where
\begin{equation}
\alpha = \left\{-1.58 \pm 0.15  E \le 1000\: {\rm {MeV}} \atop {-2.04 \pm 0.30  E \ge 1000\: {\rm {MeV}}}\right. 
\end{equation}

\noindent  The reduced $\chi^2$ for this fit is 1.17; therefore, this is not a significantly better fit.  The reason for favoring the broken power law rather than the single power law in the EGRET energy range is that the single power law is marginally inconsistent with the upper limit from OSSE.  We conservatively treat the COMPTEL results as upper limits; if the evidence for the narrow pulse were treated as a detection (which would be a flux of (6.3 $\pm$ 1.8) $\times$ 10$^{-7}$ ph cm$^{-2}$ s$^{-1}$ in the energy range 0.75--30 MeV), then a spectral break would be required in or just below the COMPTEL energy range in order to match the OSSE upper limit.  An extrapolation of the two-component spectrum back to the X-ray band is consistent with the flux seen in the 1-2 keV range, suggesting that the spectrum could be continuous across five decades in energy. 

Integrating either equation (2) or (3) gives a photon flux above 100 MeV of (1.9 $\pm$ 0.2) $\times$ 10$^{-7}$ ph cm$^{-2}$ s$^{-1}$, where the errors are statistical only. The difference between this value and the value in section 4.2 gives a measure of the systematic uncertainty that can be introduced by the two different analysis methods.  The spectrum derived here is significantly steeper than that found in the original detection of this pulsar by Fierro et al. (1993), which had a power law index of 1.18 $\pm$ 0.16 between 100 MeV and 4 GeV, based on just five broad energy bands.  The difference appears to arise from the increase in statistics, already noted by Fierro (1995).  In particular, the spectrum is now seen down to 70 MeV and up past 4 GeV with no indication of a  high-energy spectral break.  The 4-10 GeV band represents an excess of 5 photons, none of which exceed 7 GeV.  This data point does not, therefore, constrain the spectral shape, which could have either a sharp cutoff, a gradual cutoff, or no cutoff below 10 GeV.  The lack of pulsed emission in the TeV range as seen by CANGAROO (Susukita 1997) requires a steepening in the spectrum at some energy above the range detected by EGRET.

\section{Discussion}

LIGHT CURVE

The overall gamma-ray light curve for PSR B1055$-$52 differs from those of most of the other gamma-ray pulsars.  The Crab, Vela, Geminga, and PSR B1951+32 light curves are all characterized by two narrow pulses separated by 0.4$-$0.5 in phase.  PSR B1509$-$58 (detected up to 10 MeV, Kuiper et al. 1998) has a well-defined single pulse.   PSR B1055$-$52 shows two broader pulses with a phase separation of about 0.2, similar to PSR B1706$-$44.  What is common to all the pulsars seen above 100 MeV is the double pulse shape, suggestive of a hollow cone or similar geometry and consistent with the idea that these relatively young pulsars radiate primarily from the magnetospheric region associated with one magnetic pole of the neutron star (e.g. Manchester 1996). 

Comparison with the pulsar light curves at lower frequencies, Fig. 2, shows that the emission is quite complicated.  The broad  hard-X-ray pulse coincides approximately in phase but not in shape with the high-energy gamma-ray light curve, and neither of these light curves resembles that seen in soft X-rays or radio.  One component of the soft X-ray pulse is aligned with one of the radio pulses, but it has been argued based on radio polarization studies that the other radio pulse is the one that defines the closest approach to the magnetic pole (Lyne and Manchester 1988).  With pulsed emission at some wavelength seen during more than half the rotation of the neutron star, it would seem difficult to have all these components originating in one region of the magnetosphere. 

DISTANCE

The distance determined from the dispersion measure 30.1 cm$^{-3}$ pc (Taylor \& Cordes 1993) is 1.5 ($\pm$ 0.4) kpc.   Independent distance limits derived from HI absorption or other indicators are not available for this pulsar (Taylor, Manchester, and Lyne 1993).  In their analysis of the ROSAT X-ray data, \"Ogelman and Finley (1993) found that a distance of 500 pc would produce a more realistic estimate of the neutron star radius (15 km compared to 30 km for the 1.5 kpc distance estimate), although Greiveldinger et al. (1996) derived an estimate of 18$^{+15}_{-4}$ km assuming a distance of 1 kpc. Combi, Romero, \& Azc\'arate (1997) derive a distance estimate of 700 pc from a study of the extended nonthermal radio source around the pulsar.  For this work, we use the 1.5 kpc distance derived from the DM, recognizing that the pulsar may be somewhat closer. 

LUMINOSITY AND BEAMING

In terms of the observed energy flux F$_E$, the  luminosity of a pulsar is 
\begin{equation}
L_{\gamma} = 4\pi f F_E  D^2
\end{equation}
 where f is the fraction of the sky into which the pulsar 
radiates and D is the distance to the pulsar. 
This beaming fraction f is uncertain.  In a nearly-aligned rotator model, Sturner and Dermer  (1994) find a beaming 
fraction of $<$ 0.1, while an outer-gap model (Yadigaroglu and Romani 1995) suggests a value of 0.18.  In comparing the EGRET-detected 
pulsars, Thompson et al (1994) adopted a value of 1/4$\pi$.

The observed energy flux obtained by integrating eq. (3) in the range 70 MeV $-$ 10 GeV is  (1.9 $\pm$ 0.2) $\times$ 10$^{-10}$ ergs/cm$^2$ s.
For a distance of 1.5 ($\pm$ 0.4) kpc, the gamma-ray luminosity of PSR B1055$-$52 is then
(4 $\pm$ 2) $\times$ 10$^{33}$ $\times$ 4$\pi$ f ergs/s.  Unless the beaming fraction is extremely small (compared to the assumed value of 1) or the distance is less than 1 kpc, the 
observed gamma radiation represents about 6--13\% of the spin-down luminosity,
$\dot{\rm{E}}$	= 3  $\times$ 10$^{34}$ erg/s.  In light of the fact that the pulsar is not seen at TeV energies (Susukita, 1997), the spectrum must show a further steepening somewhere above the EGRET range, and the luminosity is dominated by the radiation observed in the gamma-ray band.  Extrapolating equation (3) to cover the entire range 1 keV to 30 GeV produces an energy flux and corresponding luminosity just 50\% larger than that actually measured by EGRET in the 70--10,000 MeV range.

PULSAR MODELS

Two general classes of models have been proposed for high-energy pulsars.  In polar cap models (recent examples: Daugherty and Harding 1994, 1996; Sturner and Dermer 1994; Usov and Melrose 1996; Rudak and Dyks 1998), the particle acceleration and gamma-ray production take place in the open field line region above the magnetic pole of the neutron star.  In outer gap models (recent examples:  Romani and Yadigaroglu 1995; Zhang and Cheng 1997; Wang et al. 1998), the interaction region lies in the outer magnetosphere in vacuum gaps associated with the last open field line.  Other models include a hybrid model (Kamae and Sekimoto 1995) and a Deutsch field model (Higgins and Henriksen 1997, 1998).  Romero (1998) discusses current models in light of the PSR B1055$-$52 observations.

Because all these models can be viewed as having a hollow surface geometry, a double pulse has a straightforward explanation.  The observer's line of sight cuts across the edge of the cone at two places.  Although the specific details depend on the size of the beam and its relationship to the rotation axis and the observer's line of sight, in the case of PSR B1055$-$52, one possibility is that the line of sight is closer to the edge of the cone than for the pulsars with two widely-spaced light curve peaks.  The fact that the peaks are broader for PSR B1055$-$52 is also consistent with this geometric picture, because the line of sight crosses the cone at a shallower angle. 

  In the polar cap models, a sharp turnover is expected in the few to 10 GeV energy range due to attenuation of the gamma ray flux in the magnetic field (Daugherty and Harding 1994).  The outer gap model predicts a more gradual turn-over in the same energy range (Romani 1996).  The present observations do not conflict with either model.  

\section{PSR B1055$-$52 in Comparison to Other Gamma-Ray Pulsars}

In addition to the Compton Observatory, other space- and ground-based observatories have provided a wide range of results on pulsars.  Multiwavelength energy spectra provide one useful way of comparing different pulsars across the electromagnetic spectrum.  In particular, such spectra can address such questions as the number of different emission components required.  Fig. 4 shows the broad-band energy spectra of the seven known gamma-ray pulsars.  The format is a  $\nu F_{\nu}$ or E$^2$ $\times$ Flux spectrum, showing the observed power per logarithmic energy interval.  What is shown is emission from close to the neutron star itself, either pulsed or seen as a spatially point-like source.  Although likely to be powered by the pulsar, any nebular emission is excluded.  References for this figure are given in Table 3.  An earlier version of this figure was given by Thompson (1996). 

\placetable{3}

These multiwavelength spectra have some common features:

1. The radio emission appears to be distinct from the higher-energy emission.  In most cases, the radio spectra show decreasing power at higher frequencies.  The high-energy radiation power rises from the optical to the X-ray band.  It has long been thought that the radio is a coherent process, while the high-energy radiation results from incoherent physical processes.

2. In all cases, the maximum observed energy output is in the gamma ray band.  The peak ranges from photon energies of about 100 keV for the Crab to photon energies above 10 GeV for PSR B1951+32.  This feature emphasizes that these pulsars are principally nonthermal sources with particles being accelerated to very high energies. 

2. All these spectra have a high-energy cutoff or break.  For PSR B1509$-$58, it occurs not far above 10 MeV photon energy (Kuiper et al. 1998);  for PSR B1951+32 it must lie somewhere above 10 GeV, between the highest energy EGRET point and the TeV upper limit.  As discussed above, the origin of this break can be explained in different ways by different models.

Based on their known timing ages and spectral features, these seven pulsars can be divided into two groups: the young ($\sim$ 1000 year old) pulsars, and the older pulsars.  With a timing age of about half a million years, PSR B1055$-$52 is the oldest of the gamma-ray pulsars.

\subsection{Young Pulsars}

Both the Crab and PSR B1509$-$58 have high-energy spectra that could be continuous from the optical to the high-energy gamma-ray range.  In particular, neither shows evidence of thermal emission from the neutron star surface or atmosphere (see Becker and Tr\"umper 1997, for a summary of soft X-ray properties of neutron stars), although these young neutron stars are expected to be hot ($>$ 10$^6$ K, see Page and Sarmiento 1996, for a summary).  The magnetospheric emission from accelerated particles strongly dominates the observed radiation, even in the soft X-ray band.

In the case of PSR B1509$-$58, the high-energy emission is only observed with certainty from the soft X-ray to medium gamma-ray energy ranges.  There is a candidate optical counterpart (Caraveo, Mereghetti, and Bignami 1994), but the absence of pulsations and the possibility of a chance coincidence leave some doubt that it is actually the pulsar (Chakrabarty and Kaspi 1998; Shearer et al. 1998a); hence we show the counterpart as an upper limit.  Additionally, all the points above 5 MeV (about 10$^{21}$ Hz) are upper limits, although detection by COMPTEL is now reported up to 10 MeV (Kuiper et al. 1998).  In particular, the EGRET limits (Brazier et al. 1994; Nel et al. 1996), compared with the OSSE (Matz et al. 1994) and COMPTEL (Hermsen 1997) detections show that the spectrum must bend between 10 and 100 MeV.  This spectral feature in the MeV range is unlike those seen in any of the other gamma-ray pulsars and is probably attributable to the high magnetic field of PSR B1509$-$58 (e.g. Harding, Baring, and Gonthier, 1997)

\subsection{Older Pulsars}

All five older gamma-ray pulsars share the spectral feature of having their maximum luminosity in the high-energy gamma-ray regime.  In the case of PSRs B1951+32 and B1055$-$52, the actual peak luminosity lies near or beyond the highest-energy EGRET detection of the pulsars, although the TeV limits require a turn-over in the 10  -- 300 GeV range.  The two brightest and closest of these pulsars, Vela and Geminga, show relatively sharp spectral turn-overs in the few GeV energy range.  PSR B1706$-$44 is well described by two power laws, with a spectral slope change ($\Delta\alpha \simeq$ 1) at 1 GeV.  As discussed above, PSR B1055$-$52 is consistent with also having a spectral change at 1 GeV, although smaller in magnitude than that seen in PSR B1706$-$44.

The three older gamma-ray pulsars that are the strongest X-ray sources (Vela, Geminga, and PSR B1055$-$52) all show evidence of thermal emission (as does PSR B0656+14, a possible eighth gamma-ray pulsar) consistent with emission from near the neutron star surface.  This component of the emission is clearly distinct from the nonthermal hard X-rays and gamma-rays.  Whether the hard X-ray component seen in these pulsars extends to gamma-ray energies is problematic.  In the case of Geminga, the hard component appears to extrapolate below the EGRET observations (Halpern and Wang 1997). 

\subsection{High-Energy Luminosity}

Except for the thermal peak seen in three of the pulsars of Fig. 4, the optical through gamma-ray spectra are fairly continuous, suggesting an origin in a single population of accelerated particles, though perhaps with two or more emission mechanisms. The broad-band spectra can be used to derive a high-energy luminosity, L$_{HE}$, for these pulsars, including all the observed radiation.  Integrating the observed spectra  to derive the energy flux F$_E$ is a first step, although some assumptions must be made for bands where the pulsars are not seen.  In most cases, the luminosity is dominated by the energy range around the peak in the $\nu F_{\nu}$ spectrum, as noted for PSR B1055$-$52 (so that the thermal peaks seen for Vela, Geminga, and PSR B1055$-$52 make no significant contribution).  Only in the case of the Crab is it necessary to include all the radiation from optical to high-energy gamma rays in order to estimate the luminosity.  In the cases of PSR B1509$-$58 and  
PSR B1951+32, the shape of the spectrum above the peak is unknown.  We have assumed a spectral slope change of $\Delta\alpha$ = 1.5, a value between the sharp turnover seen for Vela and Geminga and the shallower slope change for PSR B1055$-$52 and PSR B1706$-$44.  

As discussed above, there are two significant uncertainties in converting from F$_E$ into high-energy luminosity: the beaming fraction and the distance.  For the distance, we take the radio measurements and include a 25\% uncertainty, the typical value estimated by Taylor and Cordes (1993) for these pulsars.  The beaming fraction is model-dependent, but cannot exceed 1.  The fact that the high-energy pulses are typically broad suggests, but does not prove, that the beam   
 is not tiny.  We therefore adopt a value of 1/4$\pi$, assuming radiation into 1 steradian, as an intermediate value, easily scaled for comparison with models.  In most X-ray papers, luminosity is calculated assuming radiation into 4$\pi$.  If this same assumption were made for PSR B1055$-$52, the gamma rays would represent 160\% of the spin-down luminosity  $\dot E$ for the nominal distance of 1.5 kpc, or 80\% for a  
distance of 1 kpc.  The rotational energy loss of the neutron star, as the energy source, must first accelerate particles which then radiate gamma rays. It seems highly unlikely for such processes to take place with approximately 100\% efficiency.  The smaller assumed beaming fraction in this work should be taken into account when comparing with X-ray luminosities such as those in the summary of Becker and Tr\"umper (1997).  Because the distance enters the luminosity calculation as the second power, its uncertainty is likely to dominate. 

Table 4 summarizes some observed and derived properties of the seven gamma-ray pulsars, including the integrated energy flux and high-energy (optical and above) luminosities calculated here.  The efficiency $\eta$ is the ratio of the high-energy luminosity to the total spin-down luminosity.  Rudak and Dyks (1998) find similar numbers in their summary of published gamma-ray pulsar results. 
 
\placetable{4}

Fig. 5 shows the high-energy luminosity, integrated over photon energies above 1 eV under the above assumptions, as a function of the Goldreich-Julian current $\dot{\rm N}$ $\simeq$ 1.7 $\times$ 10$^{38}$$\dot{\rm P}^{1/2}$ P$^{-3/2}$ particles s$^{-1}$(Goldreich \& Julian 1969; Harding 1981) , which is also proportional to the open field line voltage V $\simeq$ 4 $\times$ 10$^{20}$ $\dot{\rm P}^{1/2}$ P$^{-3/2}$ volts ($\sim$ B/P$^2$, where B is the surface magnetic field, Ruderman and Sutherland 1975).  Both V and $\dot{\rm N}$ are proportional to $\dot{\rm E}^{1/2}$ (cf. Fig. 7 of Arons 1996 and Fig. 2 of Rudak and Dyks 1998).  Although not a perfect fit, this relationship  is a reasonable approximation extending for more than two orders of magnitude. This  proportionality would be expected if either (1) all pulsars accelerate particles to the same energy but the particle current differs from pulsar to pulsar or (2) the particle flow is constant, with different pulsars accelerating particles to different energies (Thompson et al. 1997). As noted by Arons (1996), this simple trend cannot extend to much lower values of V or $\dot{\rm N}$, because more than 100\% efficiency for conversion of spin-down luminosity would be implied.  Nevertheless, Fig. 5 shows a useful parameterization of high-energy pulsar properties with straightforward (though not unique) physical interpretations.  As noted by Goldoni and Musso (1996), other simple parameters are not well correlated with the observed properties of these pulsars.  
  The figure is similar to a pattern seen in 0.1 - 2.4 keV X-rays by Becker and Tr\"umper (1997).  The slope of the line in Fig. 5 is flatter than the one found by Becker and Tr\"umper (1997),  because the integrated luminosity is dominated by the gamma-rays, and the pulsars with smaller $\dot{\rm E}$ are also the older pulsars that have flatter energy spectra.

\section{SUMMARY}

PSR B1055$-$52 is one of at least seven spin-powered pulsars seen at gamma-ray energies.  
Observations from telescopes 
on the Compton Gamma Ray Observatory between 1991 and 1998 have 
provided new details of the gamma radiation:

1. The light curve has two peaks separated by about 0.2 in phase.  Only PSR B1706$-$44 shows a similar light curve.

2. There is no detectable unpulsed gamma radiation from the pulsar.

3. There is no evidence that the gamma-ray flux from the pulsar varies on long time scales.

4. The gamma-ray energy spectrum above 70 MeV can be represented by a  power law with photon index $-$1.73.  There may be a break in the spectrum at $\sim$1000 MeV.  

5. The maximum observable power from the pulsar is in the gamma-ray energy 
range.   

6. The observed gamma radiation represents about 6--13\% of the spin-down luminosity of the pulsar, although the unknown beaming geometry and distance uncertainty make this estimate rather uncertain. 

7. A comparison of PSR B1055$-$52 with the other gamma-ray pulsars shows that this is the oldest and most efficient in converting spin-down luminosity into high-energy radiation. 

\acknowledgments

     The EGRET team gratefully acknowledges support from the
following: Bundesministerium fur Forschung und Technologie,
Grant 50 QV 9095 (MPE authors); NASA Grant NAG5$-$1742 (HSC);
NASA Grant NAG5$-$1605 (SU); and NASA Contract NAS5$-$31210
(GAC).

\vfil
\eject

\clearpage

\figcaption[]{High-energy gamma-ray light curve for PSR B1055$-$52.  The 197 ms period is divided into 30 phase bins.  The double radio pulse has peaks with centroids at phases 0.43 (short-dash vertical line) and 0.0 (long-dash vertical line).  Top: gamma rays above 240 MeV selected from a fixed cone with radius 1.7$^{\circ}$; bottom: gamma rays above 600 MeV selected from an energy-dependent cone, as described in the text. }

\figcaption[]{Multiwavelength light curves for PSR B1055$-$52. The spectral band and intensity units are as follows: a: radio, 1520 MHz, relative intensity; b: ROSAT X-rays $<$0.5 keV, \"Ogelman and Finley (1993); c: ROSAT X-rays $>$ 0.5 keV, \"Ogelman and Finley (1993); d: OSSE gamma rays, 48--184 keV photons/s-detector; e: COMPTEL gamma rays  0.75--30 MeV; f. EGRET gamma rays  $>$ 240 MeV. Two complete cycles are shown.  The radio reference is shown by a vertical dashed line.}

\figcaption[]{Phase-averaged gamma-ray energy spectrum of PSR B1055$-$52. The fits in the energy range 70 MeV -- 10000 MeV to a  power law and a power law with a break at 1000 MeV  are described in the text, with extrapolations to lower energies shown as dotted and dashed lines respectively.  The uncertainties shown are statistical only.}

\figcaption[]{Multiwavelength energy spectra for the known gamma-ray pulsars. References for this figure are given in Table 3.}

\figcaption[]{Pulsar luminosities, integrated over photon energies above 1  eV, assuming radiation into one steradian, vs Goldreich-Julian current $\dot{\rm N}$ $\sim$ $\dot{\rm E}^{1/2}$.}

\vfil
\newpage

\begin{deluxetable}{lr}
\tablewidth{33pc}
\tablecaption{\label{1} BASIC DATA}
\tablehead{
\colhead{Parameter}           & \colhead{Value}  }
\startdata
Names  & PSR B1055$-$52,   PSR J1057$-$5226\nl

Period P&	   	0.1971  s \nl

Period Derivative $\dot{\rm{P}}$	&	5.8 $\times$ 10$^{-15}$ s/s \nl

Timing Age $\tau$		& 537,000 years\nl

Spin-down luminosity 	$\dot{\rm{E}}$	&	3.0 $\times$ 10$^{34}$ erg/s \nl

Magnetic Field B	&	1.1 $\times$ 10$^{12}$ gauss \nl

\enddata
\end{deluxetable}

\vfil
\newpage

\begin{table}
\centering
\caption{\label{2} Radio Timing Parameters for PSR B1055$-$52}
\bigskip
\begin{tabular}{lcccc}
\hline\hline
            & $T_0$ & $\nu$ & $\dot\nu$ & $\ddot\nu$ \\
Valid Dates & (MJD) & (s$^{-1}$) & ($10^{-13}$~s$^{-2}$) & 
($10^{-24}$~s$^{-3}$) \\
\hline
1991 Sep 13--1992 Oct  1 & 48704 & 5.0733041127598 & $-$1.50169 & 0.00  \\
1991 Feb  9--1994 Jan  21& 48834 & 5.0733024258637 & $-$1.50152 & 8 $\times 10^{-6}$ \\
1993 Aug 26--1995 Jan 21 & 49481 & 5.0732940338018 & $-$1.50096 & 3 $\times 10^{-5}$  \\
1995 May  6--1995 Oct 3 & 49918  & 5.0732883705741 & $-$1.50189 & 5.04  \\
1996 Feb  1--1996 Nov 12 & 50256 & 5.0732839892859 & $-$1.50195 & 3 $\times 10^{-5}$ \\  
1997 Jan 12--1998 Apr 24 & 50693 & 5.0732783214141 & $-$1.50110 & 17.2 \\ 
\hline
\end{tabular}
\end{table}

\vfil
\newpage

\begin{table}
\centering
\caption{\label{3} References for Multiwavelength Spectra}
\bigskip
\begin{tabular}{l}
\hline\hline
\\
Separate Deluxetable File\\ 
\hline
\end{tabular}
\end{table}

\vfil
\newpage

\begin{table}
\centering
\caption{\label{4} Summary Properties of the Known Gamma-Ray Pulsars}
\bigskip
\begin{tabular}{lrrccccc}
\hline\hline
 Name  & P & $\tau$ & $\dot E$ & F$_E$ & d &  L$_{HE}$  & $\eta$ \\
     & (s) & (y) & (erg s$^{-1}$) &  (erg cm$^{-2}$ s$^{-1}$) & (kpc) & (erg s$^{-1}$)
& (E $>$ 1 eV)\\
\hline
Crab & 0.033 & 1300 & 4.5 $\times$ 10$^{38}$ & 1.3 $\times$ 10$^{-8}$ & 2.0 & 5.0 $\times$ 10$^{35}$ &  0.001 \\
B1509$-$58 & 0.150 & 1500 & 1.8 $\times$ 10$^{37}$ & 8.8 $\times$ 10$^{-10}$ & 4.4 & 1.6 $\times$ 10$^{35}$ &  0.009 \\
Vela & 0.089 & 11,000 & 7.0 $\times$ 10$^{36}$ & 9.9 $\times$ 10$^{-9}$ & 0.5 & 2.4 $\times$ 10$^{34}$ &  0.003 \\
B1706$-$44& 0.102 & 17,000 & 3.4 $\times$ 10$^{36}$ & 1.3 $\times$ 10$^{-9}$ & 2.4 & 6.9 $\times$ 10$^{34}$ &  0.020 \\
B1951+32 & 0.040 & 110,000 & 3.7 $\times$ 10$^{36}$ & 4.3 $\times$ 10$^{-10}$ & 2.5 & 2.5 $\times$ 10$^{34}$ &  0.007 \\
Geminga  & 0.237 & 340,000 & 3.3 $\times$ 10$^{34}$ & 3.9 $\times$ 10$^{-9}$ & 0.16 & 9.6 $\times$ 10$^{32}$ &  0.029 \\
B1055$-$52 & 0.197 & 530,000 & 3.0 $\times$ 10$^{34}$ & 2.9 $\times$ 10$^{-10}$ & 1.5 & 6.2 $\times$ 10$^{33}$ &  0.207 \\
\hline
\end{tabular}
\end{table}

\vfil

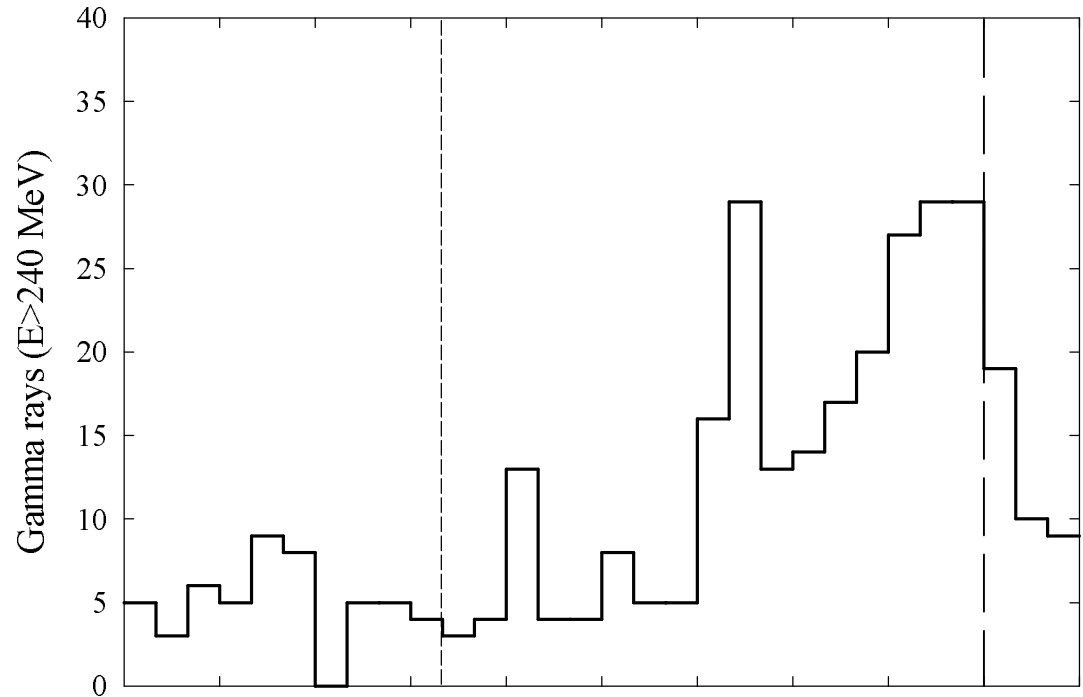
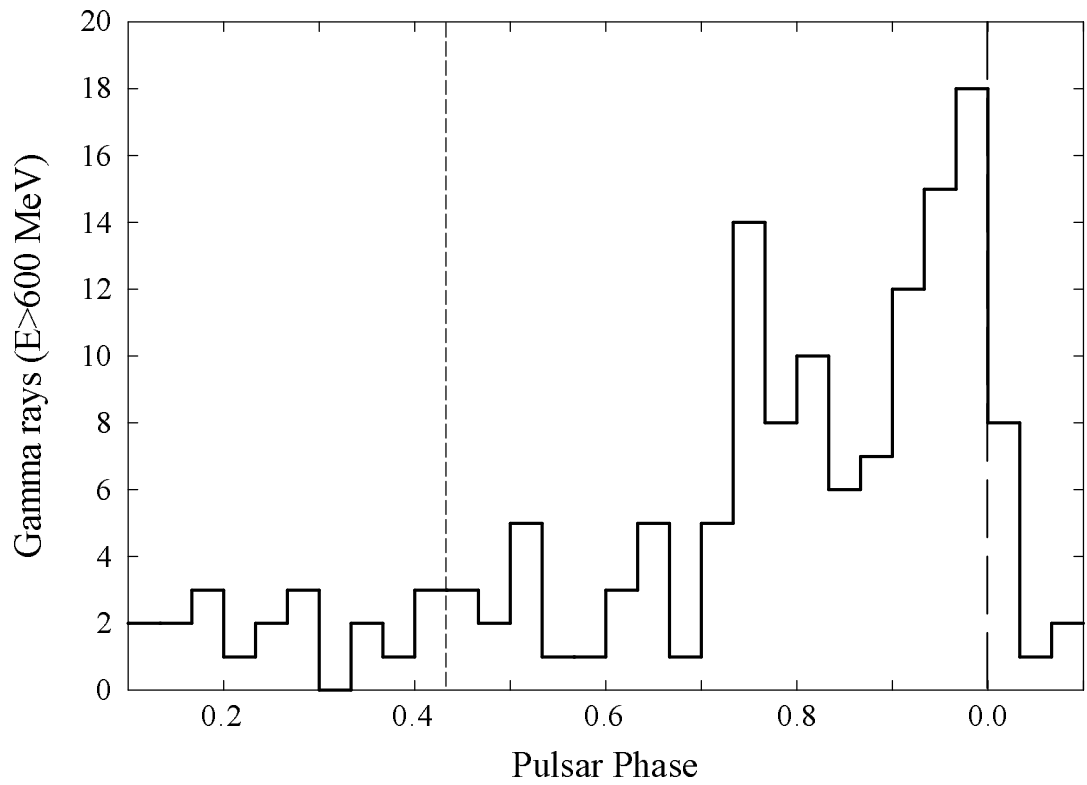
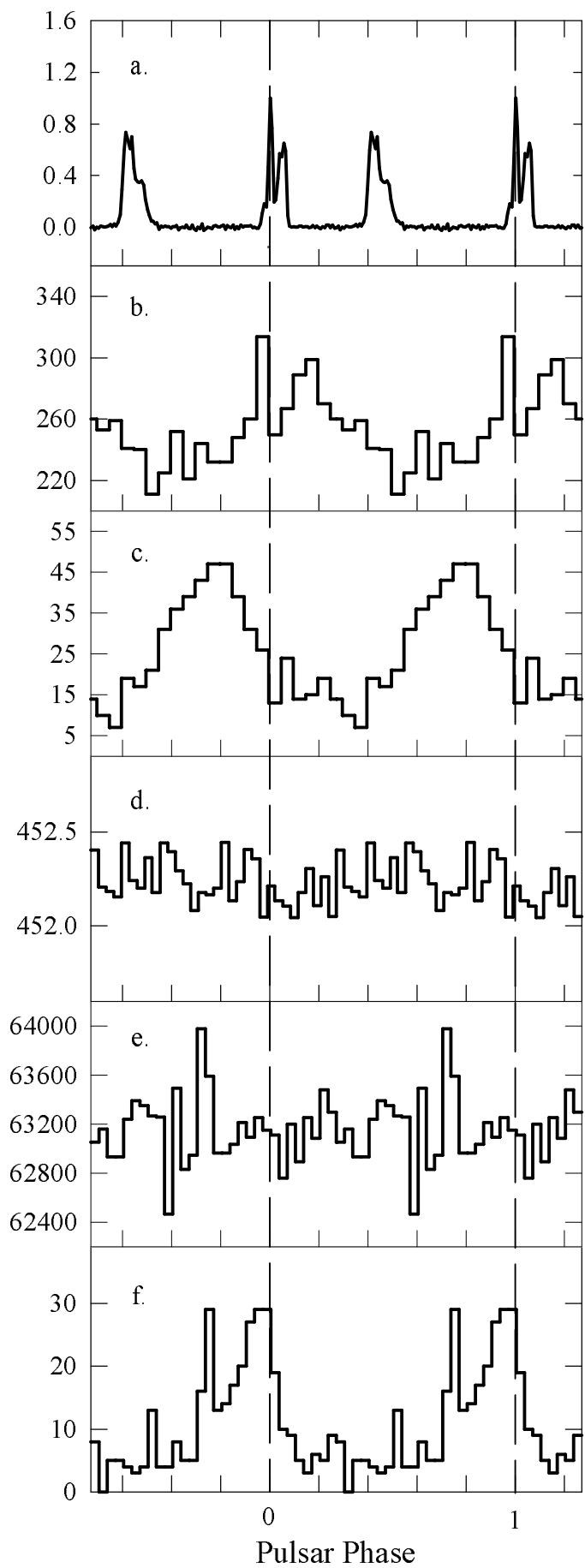
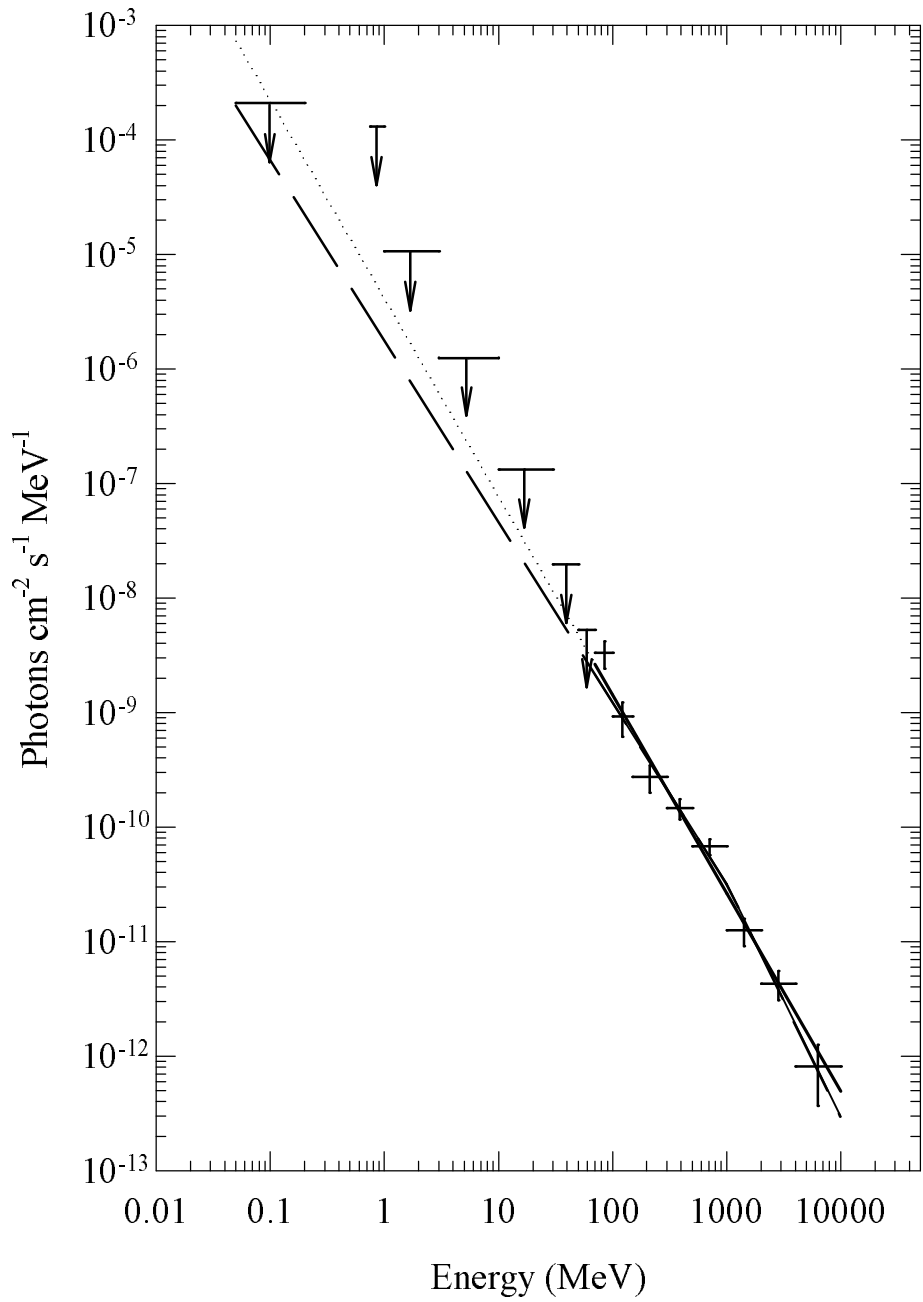
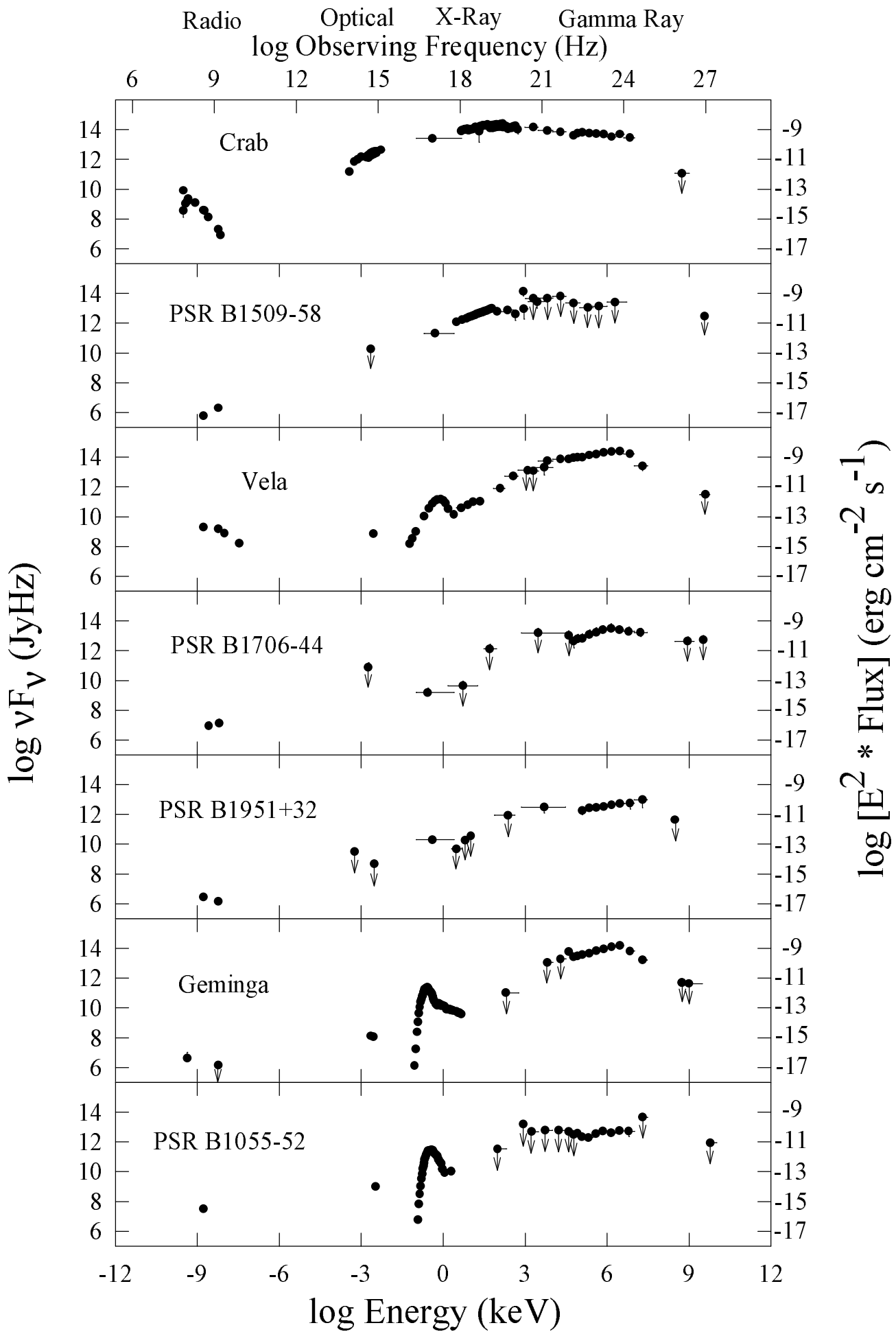
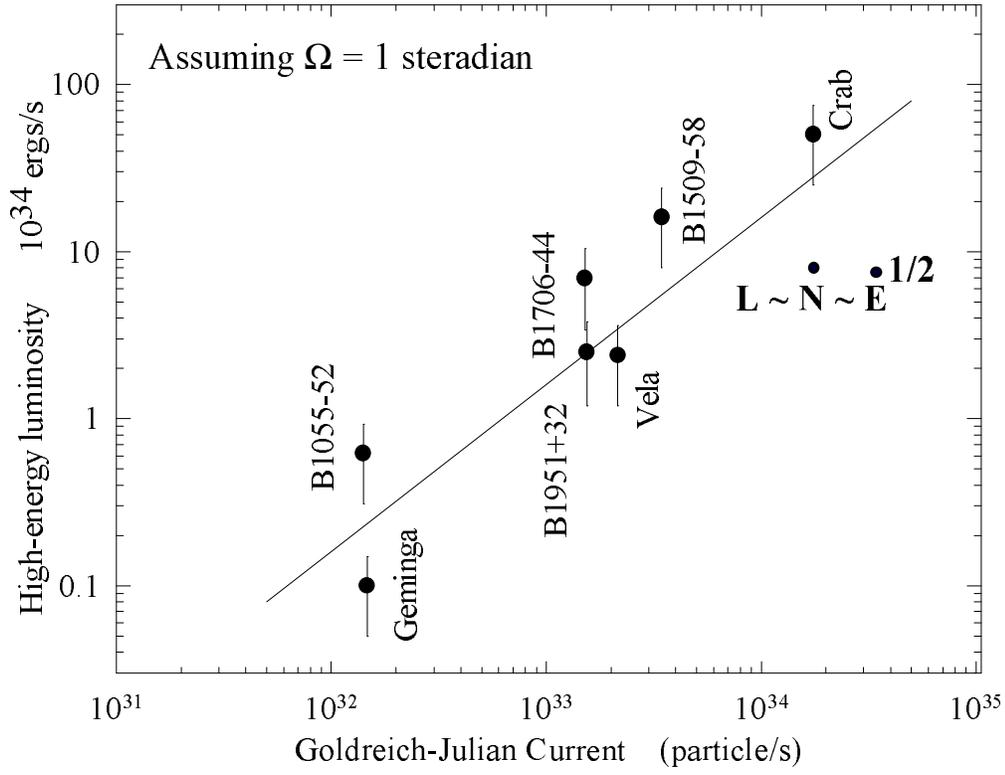
\end{document}